\documentclass[epj,draft]{svjour}

\begin{document}

\input epsf.sty

\title{Second-Order Phase Transition Induced by Deterministic
	Fluctuations in  Aperiodic Eight-State Potts 
	Models}
	\titlerunning{Aperiodic Eight-State Potts 
	Models}
	
\author{Christophe Chatelain\inst{1} \and Pierre Emmanuel Berche\inst{2} \and 
	Bertrand Berche\inst{1}\thanks{Author for correspondence 
	(berche@lps.u-nancy.fr).}}
	\authorrunning{C. Chatelain, P.E. Berche, and B. Berche}
	 
\institute{Laboratoire de Physique des
	mat\'eriaux (UMR CNRS No 7556), Universit\'e Henri Poincar\'e, Nancy 1,\\
 	F-54506 Vand\oe uvre les Nancy Cedex, France. 
 	\and 
 	Institut f\"ur Physik, Johannes Gutenberg-Universit\"at Mainz
	Staudinger Weg 7,\\ 55099 Mainz, Germany. 
	}

\date{22 June 1998}


\abstract{
 We investigate the influence of aperiodic modulations of the exchange interactions 
 between nearest-neighbour rows on the  phase transition of
the two-dimensional eight-state Potts model.  
The  systems are studied numerically through intensive Monte Carlo 
simulations using the Swendsen-Wang cluster algorithm for different aperiodic
sequences. The transition point is
 located through duality relations, and the critical behaviour is
 investigated using FSS 
 techniques at criticality. While the pure system exhibits a first-order 
 transition, we show that the deterministic fluctuations resulting from the
aperiodic coupling distribution are liable to modify drastically the physical 
properties 
in the neighbourhood of the
 transition point.
For strong enough fluctuations  of the sequence under consideration,
a second-order 
phase transition is induced. The  exponents $\beta/\nu$, $\gamma /\nu$
and $(1-\alpha)/\nu$  are obtained at the 
new fixed point and crossover effects are discussed. 
Surface properties are also studied.
}

\maketitle

\newcommand{\centre}[2]{\multicolumn{#1}{c}{#2}} 
\newcommand{\crule}[1]{\multispan{#1}{\hrulefill}} 
\def\br{\noalign{\vskip2pt\hrule height1pt\vskip2pt}} 


\section{Introduction}\label{s1}

The study of the influence of impurities on phase transitions is a quite
active field of research, motivated by  the importance of disorder in real 
experiments~\cite{Aharony78,Lubensky78,ItzyksonDrouffe89,Cardybk96}. For a disordered system to reach 
equilibrium, the time evolution should be large compared to relaxation processes
which are themselves governed by the dynamics of impurity redistributions.
In practical  experiments, such a situation can never occur in condensed
matter systems and one has to deal with quenched disorder~\cite{Brout59,Ma76}.
 
According to the Harris criterion~\cite{Harris74}, quenched bond  randomness is a 
relevant
perturbation at a second-order critical point when the specific heat exponent
$\alpha$ of the pure system is positive. 
The analogous situation when the pure system exhibits a first-order phase 
transition
was studied later.  Imry and Wortis, generalizing the Harris criterion,
argued that quenched disorder should soften the transition and could even induce 
a continuous phase 
transition~\cite{ImryWortis79}. Phenomenological renormalization-group 
studies inspired from the Imry-Ma argument for
random fields~\cite{ImryMa75}, suggest
 that in two dimensions,  an 
infinitesimal amount of randomly distributed
quenched impurities  changes the transition into a second-order 
one~\cite{Berker84,AizenmanWehr89,HuiBerker89,Berker91,Berker93}, while in larger space dimensions a finite threshold is 
necessary to produce the same effect.
The first exhaustive large-scale Monte Carlo study of  the effect of disorder 
at a  temperature-driven first-order
phase transition was performed by Chen, Ferrenberg, and 
Landau~\cite{ChenFerrenbergLandau92} who studied the two-dimensional (2D) eight-state random-bond  Potts model. 
This model is known to exhibit, in the pure version, a first-order  
transition when the
number of states $q$ is larger than  
4~\cite{Wu82}, and, the larger the value of $q$, the sharper the transition. This
property makes the Potts model a good candidate for testing the effect of
quenched bond disorder.
 Chen, Ferrenberg, and 
Landau first showed that the transition  is softened to a    
second-order phase transition in the presence of bond
randomness, and   obtained   critical
exponents very close to those of the pure 2D Ising model 
at the new critical point~\cite{ChenFerrenbergLandau95}. Since then, different 
results  obtained 
independently emerged.   
While they confirm the second-order character of the
phase transition, they conclude to a new universality 
class~\cite{CardyJacobsen97,ChatelainBerche98,Picco98}.

The essential properties of random systems are governed by disorder 
fluctuations usually described by normally distributed random variables. All physical quantities depend on the configuration of disorder,
and the study of the influence of randomness requires an
average over disorder realizations. 
Among the systems where the presence of fluctuations is also of primary 
importance,
aperiodic systems have been of considerable interest since the
 discovery of quasicrystals~\cite{ShechtmanBlechGratiasCahn84}. Quasiperiodic 
 or aperiodic distributions of couplings
 strengths appear as an alternative to quenched bond randomness, albeit
 built in a deterministic way, making any configurational average
useless. Their critical 
properties have  been intensively studied, especially in the Ising model (for 
a review, see, e.g., 
ref.~\cite{GrimmBaake96}). 
The characteristic length scale in a critical system is given by the 
correlation 
length and, as in 
 the Harris criterion for random systems, the
 coupling fluctuations  on  this scale determines the 
critical behaviour.
An aperiodic perturbation can thus be relevant, marginal or irrelevant, 
depending on the
sign of a crossover exponent involving both the correlation length exponent $\nu$ of 
the
unperturbed system  and the wandering exponent $\omega$ which governs the 
size-dependence
of the fluctuations of the aperiodic couplings~\cite{Luck93}. In the light
of this criterion, the results obtained in early papers, mainly concentrated on
the Fibonacci and the Thue-Morse 
sequences (see e.g.
Refs.~\cite{Igloi88,DoriaSatija88,Tracy88,Benza89,DoriaNoriSatija89}) 
found a consistent
explanation,
 since, resulting from 
the bounded character of fluctuations, a critical
behaviour which belongs to the pure model universality class  
 was found  in 
two dimensions.

In the last years, much progress have been made in the understanding of the
properties of marginal and relevant aperiodically perturbed systems. Exact 
results for the 
2D layered Ising model and the quantum Ising chain   have  been
obtained with irrelevant, marginal and relevant aperiodic
perturbations~\cite{TurbanIgloiBerche94,IgloiTurban94,KarevskiPalagyiTurban95,BercheBerche97b}.
The critical behaviour is in agreement with Luck's criterion, leading to 
essential singularities
or first-order surface transition when the perturbation is relevant and power 
laws with
continuously  varying exponents in the marginal situation with logarithmically 
diverging
fluctuations.  A strongly anisotropic behaviour has been recognized in this 
latter 
situation~\cite{BercheBercheTurban96,IgloiLajko96,IgloiTurbanKarevskiSzalma97}.

The effect of quasiperiodic and aperiodic distributions of exchange couplings
at first-order phase transitions has only recently been investigated.
It was shown
that the transition remains first-order for the eight-state Potts model on a
quasiperiodic tiling~\cite{LedueLandauTeillet97}, while a finite-size 
scaling study using Monte Carlo
simulations has shown strong evidences in favor of a  
second-order phase transition
for the ``Paper-Folding'' aperiodic perturbation~\cite{BercheChatelainBerche98}.
In the present paper, we report an extensive Monte Carlo study of the influence
of aperiodic modulations of the coupling strengths on the nature of the phase
transition in the 2D eight-state Potts model. We are interested in
both bulk and surface properties, and several aperiodic sequences are 
considered.

The paper is organized as follows: In Section 2, after a summary of the essential
properties of aperiodic sequences and a presentation of the layered structure 
of the system, the   critical point of the models is exactly located
through duality. A qualitative description of the phase transition is given
in Section 3 from a numerical study of the temperature dependence of some 
physical quantities. Eventually Section 4 contains the results of a Finite-Size
Scaling (FSS) analysis.


\section{Layered aperiodic structure and details of the Monte Carlo simulations}
\label{s2}

The Thue-Morse sequence is an example of aperiodic succession of digits 
$f_k=0$ or $1$
 leading
to bounded fluctuations. It may be defined as a two digits 
substitution
sequence which follows from the inflation rule 
\begin{equation}
0\rightarrow S(0)=01,\quad
1\rightarrow S(1)=10,
\label{eq-1}\end{equation} 
leading, by iterated application of the rule on the initial word $0$, to 
successive words of
increasing lengths: $\{f_k\}=0\ 1\ 1\ 0\ 1\ 0\ 0\ 1\dots$ 
It is  well known that most of the properties of such a sequence can be
characterized by a substitution matrix whose elements $M_{ij}$ are given by the
number $n_i^{S(j)}$ of occurrences of digits  $i$ in the substitution
$S(j)$~\cite{Queffelec87}.
The largest eigenvalue of the
substitution matrix
 is   related to the length of the
sequence after $n$ iterations, $L_n\sim\Lambda_1^n$, while the second 
eigenvalue
$\Lambda_2$ governs the behaviour of the cumulated deviation
from the asymptotic density 
$\rho_\infty=\bar f_k$: 
\begin{equation}
\sum_{k=1}^L(f_k-\rho_\infty  )\sim\mid\Lambda_2\mid^n\sim 
(\Lambda_1^\omega)^n,
\label{eq-4}\end{equation} 
where  the wandering exponent is defined by:
\begin{equation}
\omega={\ln\mid\Lambda_2\mid\over\ln\Lambda_1}.
\label{eq-5}\end{equation}

\begin{figure}[ht]
\epsfxsize=7cm
\begin{center}
\vglue-0cm
\mbox{\epsfbox{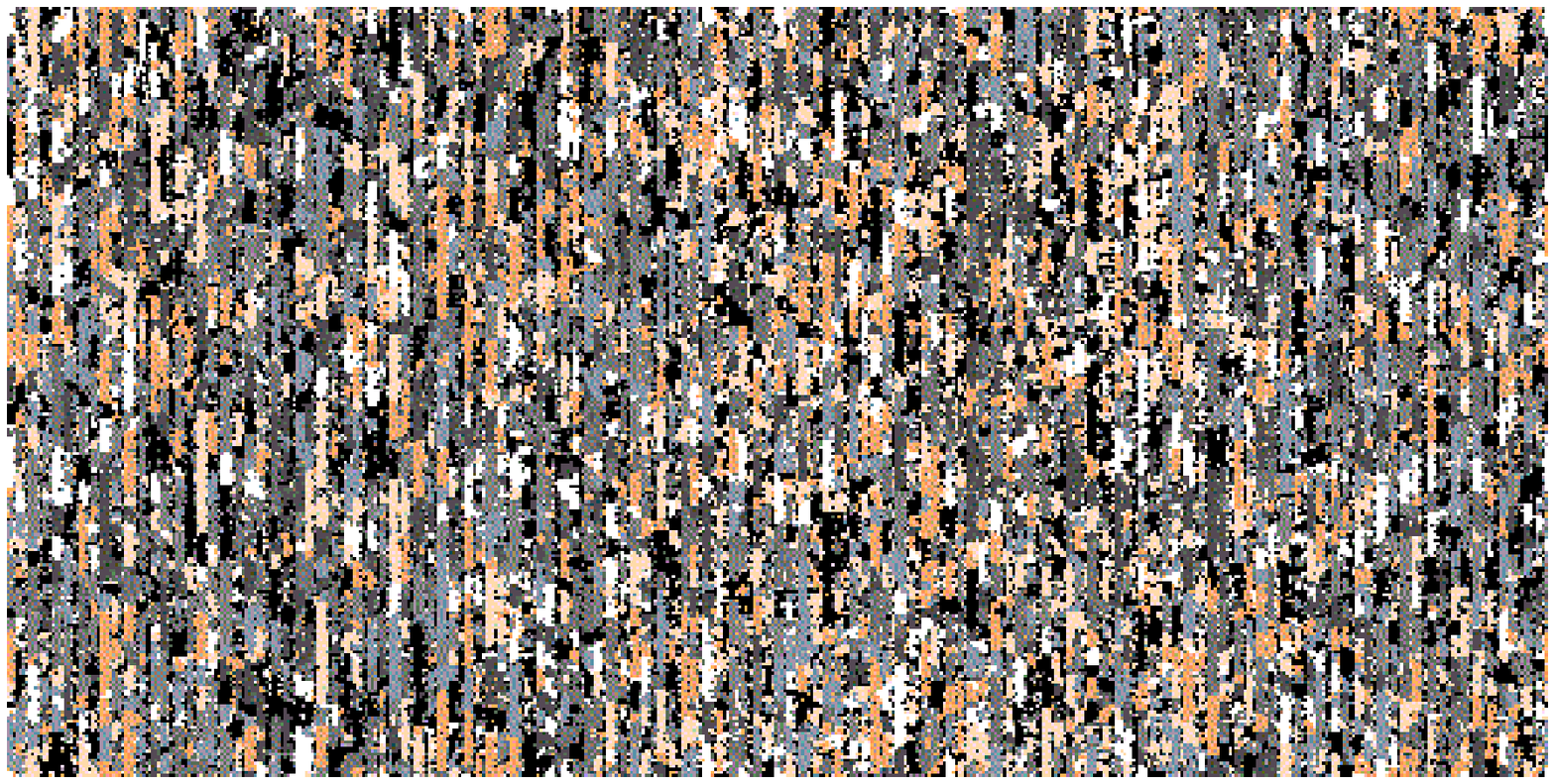}}
\vskip 3mm\epsfxsize=7cm
\mbox{\epsfbox{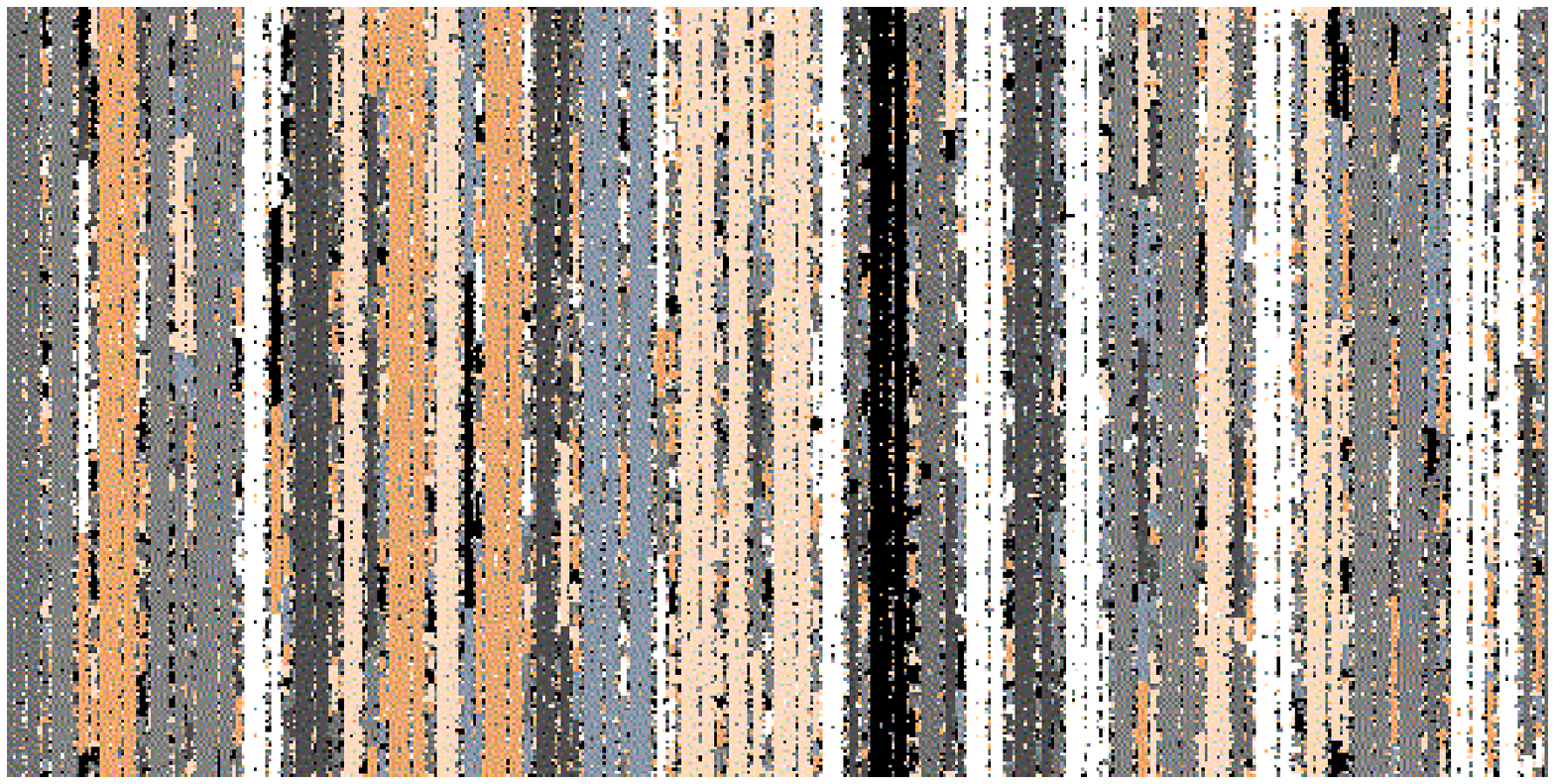}}
\vskip 3mm\epsfxsize=7cm
\mbox{\epsfbox{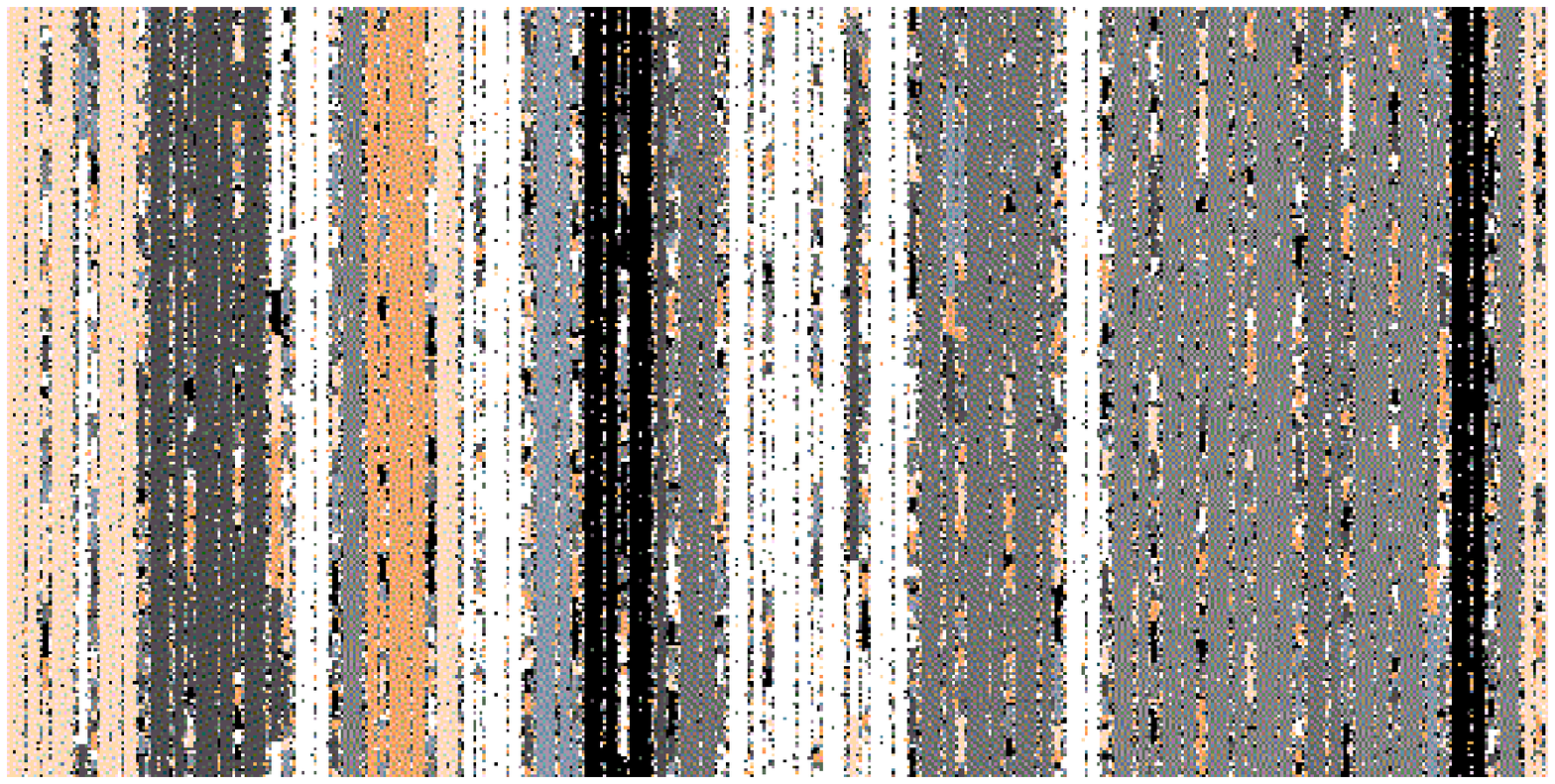}}
\vskip 3mm\epsfxsize=7cm
\mbox{\epsfbox{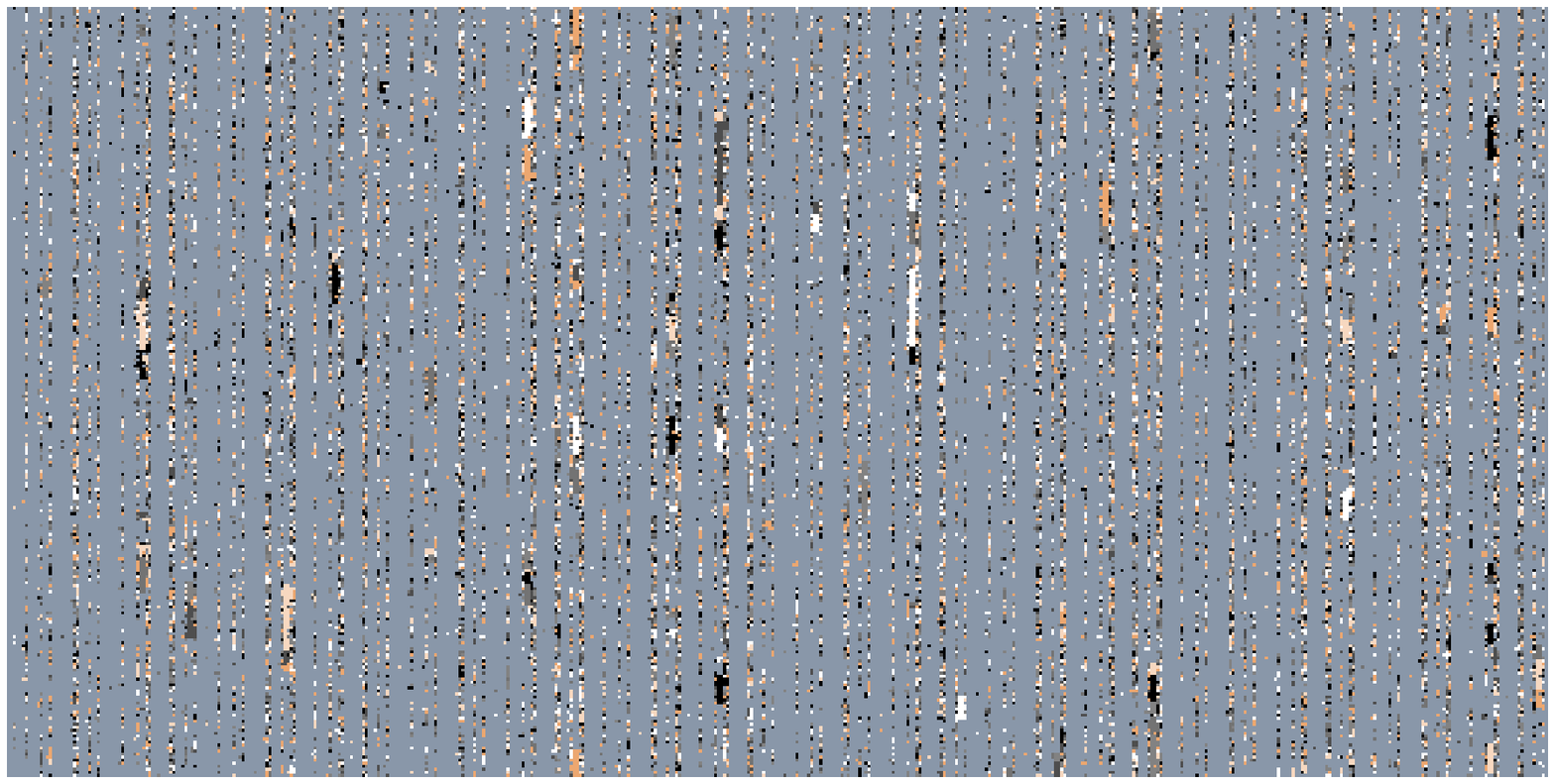}}
\end{center}
\caption{Typical Monte Carlo configurations
(system of size $256\times 512$) in the
high-temperature phase ($K_0=0.3$), in the
neighbourhood of the critical point ($K_0=0.5$ and $K_0=0.6$), 
and in the low-temperature phase  ($K_0=0.7$). 
The layered structure of the system is 
clearly
visible.}
\label{fig:MCconf}
\end{figure}

The spin system considered in the following is a layered two-dimensional 
8-state Potts model.
The Hamiltonian of the system with aperiodic interactions can be written
\begin{equation}
-\beta {\cal H}=\sum_{(i,j)}K_{ij}\delta_{\sigma_i,\sigma_j} ,
\label{eq-hamPotts}
\end{equation}
where the spins $\sigma_i$, located at sites $i$,  
can take the values $\sigma=1,2,\dots,q$,
 the sum goes over nearest-neighbour pairs, and
 the coupling 
strengths are allowed to take two different values $K_0=K$ and $K_1=Kr$. They
are distributed according to a layered structure {\it i.e.} the distribution
is translation  invariant in one lattice direction, and follows
the aperiodic modulation $\{ f_k\}$   in the other 
direction: In layer $k$, both horizontal and vertical couplings take the 
same value $Kr^{f_k}$. This layered structure is reminiscent in the shape of the
correlated clusters
obtained by Monte Carlo simulations (Fig.~\ref{fig:MCconf}).

\begin{figure}[ht]
\epsfxsize=7cm
\begin{center}
\vglue-0cm
\mbox{\epsfbox{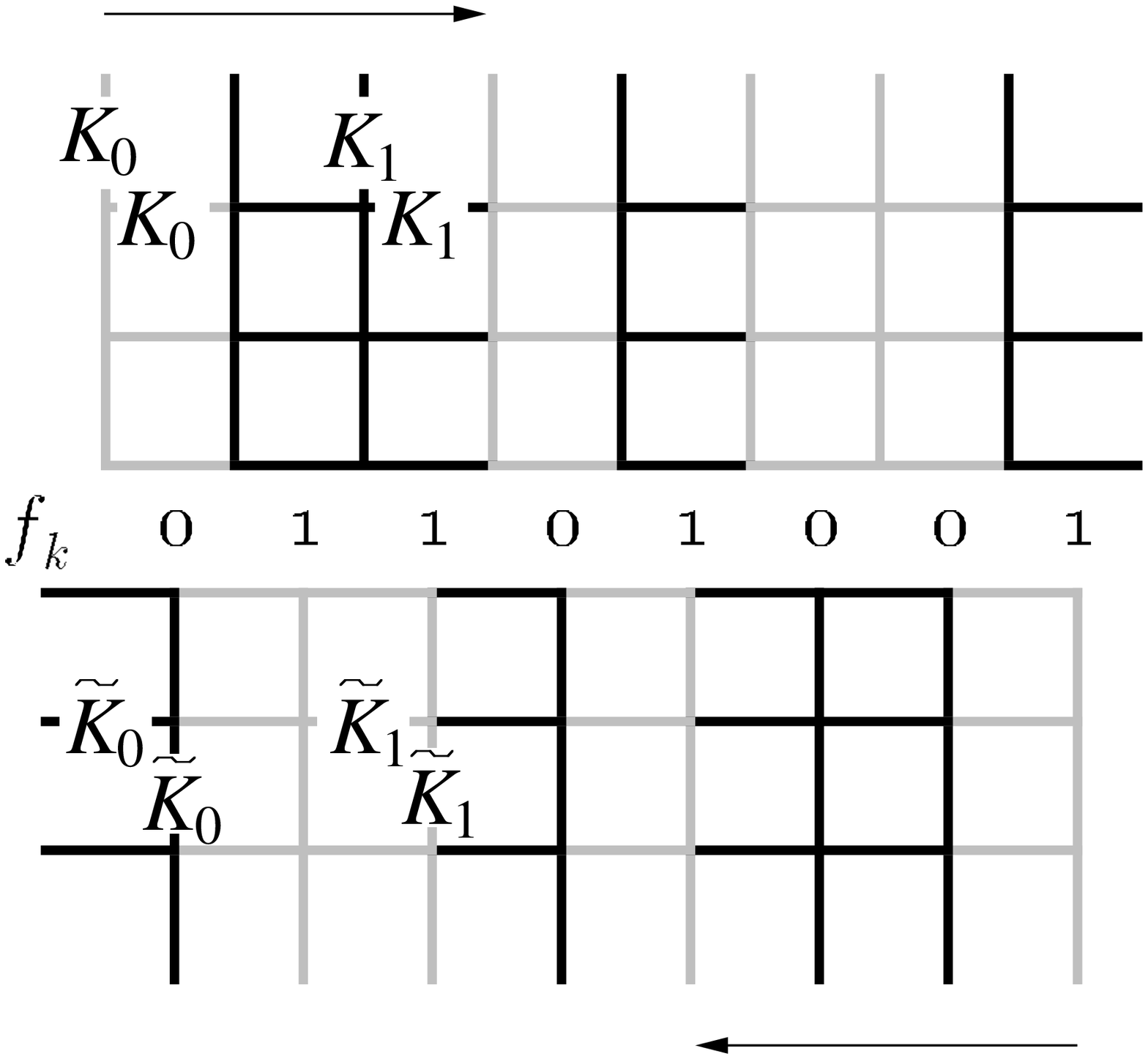}}
\vskip 3mm\epsfxsize=7cm
\mbox{\epsfbox{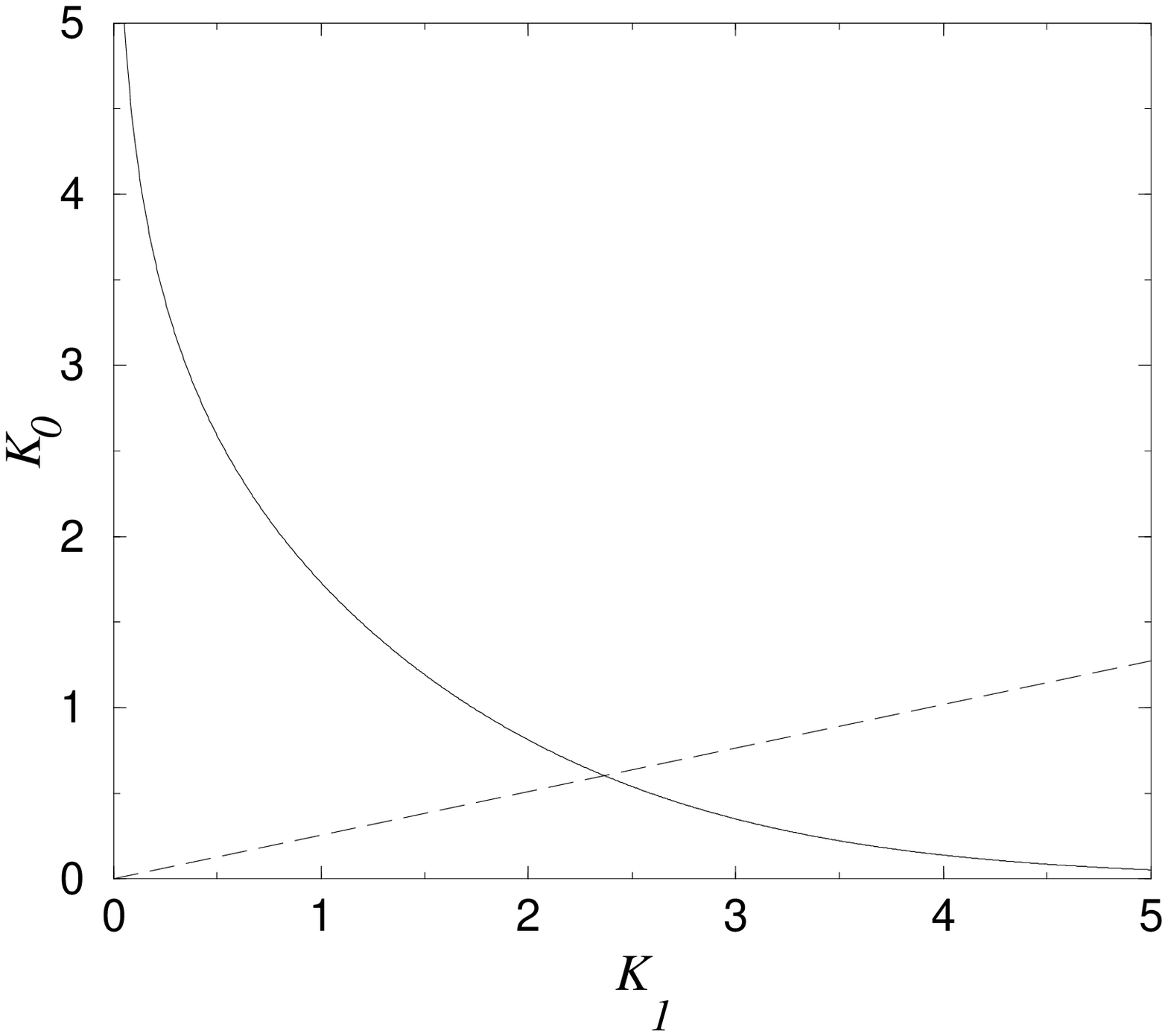}}
\end{center}
\caption{Layered structure of the system and its transformation
under duality (above). The critical line in the coupling space is also shown 
(below).}
\label{fig:dual}
\end{figure}

Particular  choices of coupling distribution make it possible to determine
exactly the critical point by duality arguments~\cite{BercheChatelainBerche98}. 
Consider a system of $N$ layers with a
distribution $\{f_k\}$, made from a succession of vertical-horizontal (V-H) 
bonds
when read from left to right (Fig.~\ref{fig:dual}), and let us write its 
singular free energy density
$f_s(K_0,K_1;\{f_k\})$.
Under a duality transformation, the strong and weak couplings $K_i$ are,
respectively, replaced by 
weak and strong dual couplings $\tilde K_i$, where 
${\rm e}^{\tilde K_i}-1=q/({\rm e}^{K_i}-1)$. Since a vertical bond on 
the original
lattice becomes horizontal on the dual system, the same V-H bond configuration 
is 
recovered for the transformed system when the distribution is read from 
right to left, and one
gets the same type of system, but a reverse distribution 
$\{f_{L+1-k}\}$.
The free energies of the two systems are equal: $f_s(K_0,K_1;\{f_k\})=
f_s(\tilde K_0,\tilde K_1;\{f_{L+1-k} \})$.
The sequences considered here have the  property that the reverse
 distribution
corresponds to the original one after  exchange of perturbed and unperturbed
couplings $K_1\leftrightarrow K_0$: 
\begin{equation}
f_s(\tilde K_0,\tilde K_1;\{f_{L+1-k} \})=
f_s(\tilde K_1,\tilde K_0;\{f_k\}).\label{eq-dual}
\end{equation}
The system being thus self-dual,
the critical point, if  unique,  is exactly given by the critical line
$(K_0)_c=(\tilde K_1)_c$
of the usual anisotropic model~\cite{Fisch78,KinzelDomany81}:
\begin{equation}
({\rm e}^{K_c}-1)({\rm e}^{K_cr}-1)=q. 
\label{eq-Kc}
\end{equation}
One should mention  that the required symmetry property of the sequences
possibly holds after omitting
the last digit which simply introduces an irrelevant surface effect.

The fluctuations of the coupling strengths per bond at the correlation length scale $\xi$
induce a thermal perturbation $\langle\delta t\rangle\sim t^{-\nu(\omega -1)}$,
which has to be compared to the deviation from the critical point $t$. The
resulting perturbation has a crossover exponent $\Phi=1+\nu(\omega -1)$ and
is relevant when $\Phi> 0$. This criterion, first obtained by 
Luck~\cite{Luck93}, founds its justification at a critical point, {\it i.e.} when the 
correlation length of the pure system diverges as the transition point is 
approached. The question of its application at first-order phase transitions
is not yet clear. The purpose of this paper is to report some results in this
situation.
The eight-state Potts model has the advantage of undergoing a strong first-order
transition in
the pure case, and has been already intensively studied by several techniques
with  random 
bonds~\cite{ChenFerrenbergLandau95,CardyJacobsen97,ChatelainBerche98}.

In the following, we consider three different aperiodic sequences and a
 periodic system (PS) with the regular
succession of couplings $K_1$, $K_0$, $K_1$, $K_0,\dots$,  in which 
the transition is surely first-order.
This system constitutes
a reference for the first-order type behaviour and presents the advantage
of having the same value for the  critical coupling (at fixed $r$)
 than the aperiodic sequences
considered. The substitution rules for the
aperiodic sequences studied in the
paper are 
given in Table~\ref{tab:table1}.
\begin{table}
\small
\caption{Substitution rules for the aperiodic sequences considered in the
text.}
\begin{center}
\vglue0mm
\begin{tabular}{@{}*{5}{l}}
\br
\centre{2}{\rule{0mm}{1mm}Sequence}  & substitutions &   wandering\\
&&& exponent\\
\hline
{\it\rule{0mm}{3mm} Thue-Morse}& {\rm (TM):}& $0\to 01,$ & $\omega_{\rm TM}=-\infty$ \\
&& $1\to 10.$ &\\
{\it Paper-Folding}& {\rm (PF):}& $00\to 1000,$ & $\omega_{\rm PF}=0$ \\
&& $01\to 1001,$ &\\
&& $10\to 1100,$ &\\
&& $11\to 1101.$ &\\
{\it Three-Folding}& {\rm (TF):}& $0\to 010,$ & $\omega_{\rm TF}=0$ \\
&& $1\to 011.$ &\\
\br
\end{tabular}
\end{center}\label{tab:table1}
\end{table}
Details on their
properties can be 
found in ref.~\cite{TurbanIgloiBerche94,BercheBercheTurban96}.
We first performed preliminary runs (to be presented in the next 
section) for different values of the temperature, and then intensive Monte 
Carlo simulations at criticality on two-dimensional square 
lattices of sizes $L\times 4L$ (PF, TM, PS) or $L\times 3L$ (TF),
using the Swendsen-Wang cluster algorithm~\cite{SwendsenWang87} .
This technique is known to
be very efficient to study second-order phase transitions, since 
it is less affected
by the critical slowing down than conventional Metropolis algorithm. At first-order 
phase transition points, other methods can be used to improve 
the effectiveness,  but we favoured the use of the same algorithm to
study the different regimes, increasing the number of Monte Carlo iterations
when necessary in order to obtain reliable results.
The multi-spin coding technique has also been 
used to speed up the simulations~\cite{CreutzJacobsRebbi81}.
The geometry $L\times pL$ allows a sufficiently large number of rows in
order to explore the aperiodic structure at long enough length scales and the
value of $p$ has been chosen with respect to the symmetries of the sequences.
The boundary conditions are periodic in the short direction ($L$) and
free or fixed in the long direction ($pL$).\footnote[1]{
The details of the simulations at the critical point (sizes, boundary 
conditions, autocorrelation time,  $\#$ of 
MC iterations) are given in Table~\ref{tab:table2}.}

\begin{table}
\small
\caption{Details of the typical parameters used in the 
Finite-Size Scaling 
Monte Carlo 
simulations$^{a}$
and energy auto-correlation time in the case $r=5$. 
Sizes $L=2^n$ or $3^n$ between the limits  indicated in the table have been 
used.}\begin{center}
\vglue0mm
\begin{tabular}{@{}*{7}{l}}
\br 
\rule{0mm}{1mm}Sequence  & size$^{(b)}$ &   {$\#$ MCS/spin$^{(c)}$} & \centre{2}{$\tau_E$}\\
                   &      &                           &   \crule{2}\\
                   &      &                         &  min. & max.   \\
\hline
PS   & 8 to 256  & $2.10^5$ to $3.10^6$   & 7.3 & 651.1 \\
TM   & 8 to 2048 & $1.10^5$ to $4. 10^6$   & 6.7 & 141.4 \\
PF   & 8 to 2048 & $3.10^5$ to $4.10^5$   & 6.7 & 8.8 \\
TF   & 3 to 2187 & $2.10^5$ to $2.5\ 10^5$   & 4.5 & 18.2 \\
\br
\end{tabular}
\\
\medskip\end{center}
{\footnotesize   $^a$5000 iterations (in MCS/spin) have been 
discarded.\hfill\break
$^b$The values indicated  correspond to the largest size $pL$
(horizontal direction) with $p=4$ (PS, TM, PF) and
$p=3$ (TF).\hfill\break
$^c$The same numbers of MC iterations have been used for the two types of 
boundary conditions (free and fixed in the horizontal direction).\hfill\break
}\label{tab:table2}
\end{table}

\begin{figure}[ht]
\epsfxsize=8cm
\begin{center}
\vglue-4cm
\mbox{\epsfbox{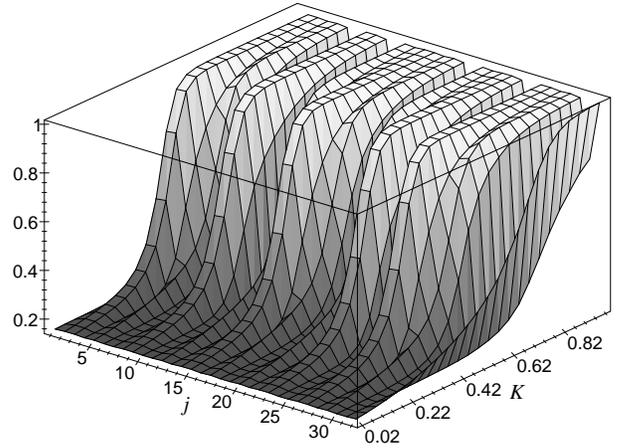}}\end{center}
\caption{Temperature dependence of the order parameter 
profile in the case of the Thue-Morse sequence.}
\label{fig:profaim}
\end{figure}

With these boundary conditions,  translational invariance holds in the vertical direction, and a local 
order parameter is
defined by the majority orientation of the spins at column 
$j$~\cite{ChallaLandauBinder86}:
\begin{equation}
m(j)=\frac{q\rho_{\rm max}(j)-1}{q-1},\quad M_j\equiv\langle m(j)\rangle.
\label{eq-mj}
\end{equation}
Here, $\rho_{\rm max}(j)={\rm max}_\sigma[\rho_\sigma(j)]$, where 
$\rho_\sigma(j)$ is the density of spins in the state $\sigma$ at column $j$ and 
$\langle\dots\rangle$ denotes the thermal average over the Monte Carlo 
iterations. The systems under consideration are highly inhomogeneous, as it can be seen
in Fig.~\ref{fig:profaim}, so,  in order to
reduce fluctuations, we studied  average quantities, e.g.
\begin{equation}
m=\frac{q\rho_{\rm max}-1}{q-1},\quad M\equiv\langle m\rangle,
\label{eq-meanmag}
\end{equation}  
where $\rho_{\rm max}$ has the same meaning as above, over the whole system and 
is not restricted to a given row. 
The susceptibility is obtained as usually via the fluctuations of magnetization
$\chi=KpL^2(\langle m^2\rangle-\langle m\rangle^2)$, and we also computed the
energy density
\begin{equation}
E=\frac{1}{2KpL^2}\left\langle\sum_{(i,j)}K_{ij}\delta_{\sigma_i,\sigma_j}
\right\rangle,
\label{eq-en-density}
\end{equation}  
where the prefactor ensures a normalization to 1.
Local properties at the surface have also been calculated, e.g. 
$M_1= \langle m(1)\rangle$.


\section{Off-critical point behaviour}\label{qual}

\begin{figure}[ht]
\epsfxsize=7.5cm
\begin{center}
\mbox{\epsfbox{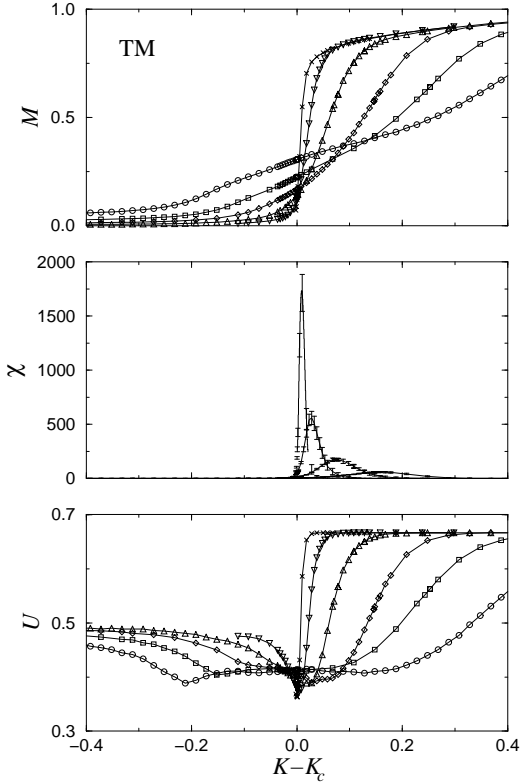}}\end{center}
\caption{Temperature dependence of the average magnetization, susceptibility, and magnetization
 cumulant for the TM sequence ($r=5$). The different symbols correspond to
simulations of systems of sizes $8\times 16$ ($\circ$) to $256\times 512$ 
($\times$).}
\label{fig:TM_tot_K}  
\end{figure}
\begin{figure}[ht]
\epsfxsize=7.5cm
\begin{center}
\mbox{\epsfbox{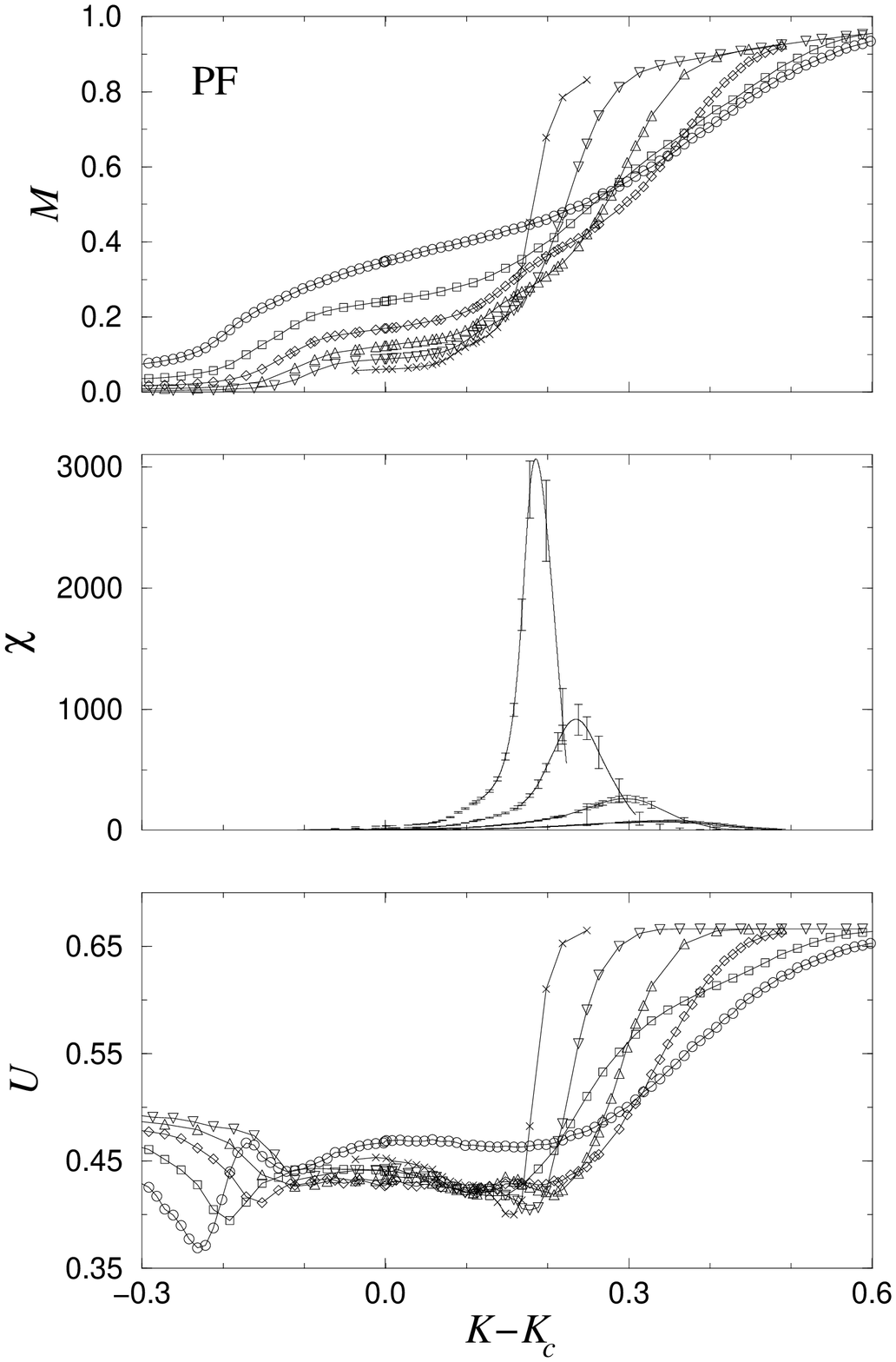}}\end{center}
\caption{Temperature dependence of the average magnetization, susceptibility, and magnetization
 cumulant for the PF sequence ($r=5$). The different symbols correspond to
simulations of systems of sizes $8\times 16$ ($\circ$) to $256\times 512$
($\times$).}
\label{fig:PF_tot_K}  
\end{figure}

In this section, we  give a qualitative description of the order of the phase 
transition. For this purpose, we performed preliminary MC simulations over a range of values
of $K$ for a system of size $L\times 2L$ ($L$ from 8 to 256, TM and PF 
sequences), and determined the temperature dependence of the physical quantities.
The behaviour of the average magnetization, susceptibility, and 
Binder cumulant of the magnetization for the TM and PF sequences at $r=5$ are shown in 
Fig.~\ref{fig:TM_tot_K} and Fig.~\ref{fig:PF_tot_K}, 
respectively.

\begin{figure}[ht]
\epsfxsize=8cm
\begin{center}
\mbox{\epsfbox{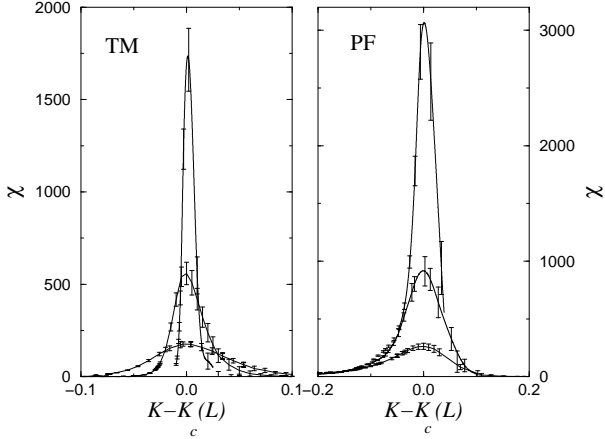}}\end{center}
\caption{Rescaled susceptibility with respect to the
size-dependent critical coupling $K_c(L)$ for TM and PF sequences at $r=5$. Three
sizes ($64\times 128$, $128\times 256$ and
  $256\times 512$) have been used. In the case of TM, one observes a behaviour which
  approaches a $\delta$-function, while a power-law is obtained for PF.}
\label{fig:Chi_TM-PF}  
\end{figure}

As well known in Monte Carlo simulations, it is difficult to observe, in the
numerical data, a jump of
the order parameter at the transition point of a first-order phase transition, 
and similarly the $\delta$-like behaviour of the susceptibility cannot easily be
distinguished from a pure power-law, so
we report here also the results for the magnetization 
cumulant~\cite{Vollmayretal93}. 
As a consequence
of the highly inhomogeneous systems under consideration, these cumulants exhibit
a quite complicated structure, but their behaviour already gives an idea of the 
nature of the transition for the two sequences.
One can indeed observe in Fig.~\ref{fig:TM_tot_K} (TM) that a narrow well appears
in the vicinity of the transition point 
and becomes deeper as the system
size increases. This should be the signature of a first-order phase transition,
while in Fig.~\ref{fig:PF_tot_K} (PF) there is no analogous significant trend. The direct comparison
between the two sequences also shows that the variation of the magnetization and
of the susceptibility close to the critical coupling is sharper for TM than PF.

A criterion to analyse the order of phase transitions, from the observation
of the way the non-analytic behaviour develops as the critical point is 
approached, 
has  recently been
proposed~\cite{Meyer-OrtmannsReisz98}. Singular quantities, like the 
susceptibility,
are response functions with diverging non-analytic behaviour in the thermodynamic limit. In 
the scaling region, there exists a certain interval where such functions are
decreasing at first-order transitions, 
leading
to crossings of the rescaled curves which evolve towards a $\delta$-like
behaviour, while they increase 
 with the size of the system in the neighbourhood of $K_c$ at 
second-order
transitions (power-law behaviour). This is illustrated in Fig.~\ref{fig:Chi_TM-PF} 
where the 
case of Thue-Morse sequence
belongs to the first situation, while Paper-Folding corresponds to the 
second one. For this latter perturbation, typical Monte Carlo
simulations are shown in Fig.~\ref{fig:MCconf}.
In the disordered phase, the layered structure of the correlated clusters
becomes appearent as the system approaches the transition point, where these
clusters begin to grow in the perpendicular direction, leading to a second
order critical point in the thermodynamic limit.


Similar qualitative observations were reported in 
Ref.~\cite{BercheChatelainBerche98}, where temperature-dependent 
effective exponents, for average magnetization and susceptibility, were
computed by comparing the data at  two different
sizes $L$ and $L'=L/2$:
 In the case of the magnetization for example, assuming 
a scaling form 
 $M_L(t)=L^{-\beta/\nu}{\cal M}(Lt^\nu)$,
 where $t=\mid\! K-K_c\!\mid$ and ${\cal M}(x)$ is a scaling function, 
  the quantity
 \begin{equation}
{X}_L(t)=\frac{\ln M_L/M_{L'}}{\ln L/L'}.
 \label{eq-X}
 \end{equation}
 expanded in powers of $Lt^\nu$ close to $K_c$ leads to
 \begin{equation}
 {X}_L(t)\simeq -\frac{\beta}{\nu}+\frac{Lt^\nu}{2\ln 2}
 \frac{{\cal M}'(Lt^\nu)}{{\cal M}(Lt^\nu)}+O(L^2t^{2\nu}) ,
 \label{eq-effective}
 \end{equation}
which defines an effective exponent which  evolves towards $-\frac{\beta}{\nu}$
as the critical point is approached and in the 
thermodynamic
 limit. 
 In the case of the TM sequence, the successive estimates of  
 $\frac{\beta}{\nu}=d-y_h$ and 
 $\frac{\gamma}{\nu}=2y_h-d$
 evolve 
 towards the values
 0 and 2. This is  characteristic of a first-order phase transition, since the scaling
dimensions associated to the temperature and magnetic field, $y_t$ and $y_h$,
respectively,  take a special value equal to the dimension $d$ of the 
system~\cite{FisherBerker82}.
  In the case of the PF sequence,
 the behaviour is drastically different, and these effective exponents 
 evolve towards non trivial values around 0.5 and 1.

\section{Finite-Size Scaling}\label{s3}

\subsection{Dynamical exponent}

\begin{figure}[h]
\epsfxsize=9.5cm
\begin{center}
\mbox{\epsfbox{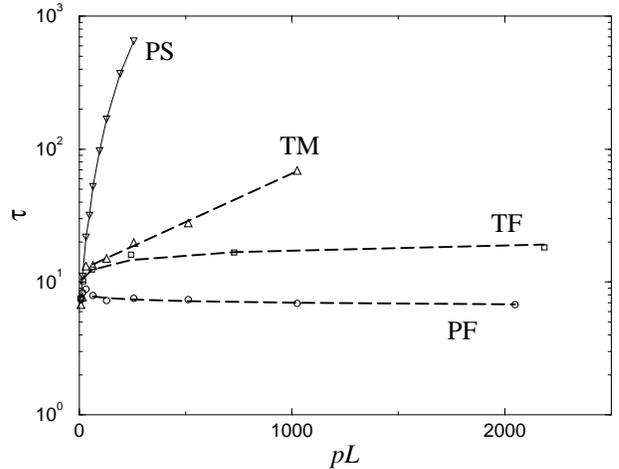}}\end{center}
\caption{Energy autocorrelation time
$\tau$ at $K_c$ ($r=5$). For TM, the dashed line is a fit to an 
exponential behaviour, while it is a power-law fit for PF ($z\sim -0.04$) 
and TF ($z\sim 0.12$) sequences. The data corresponding to the periodic system have not been fitted.}
\label{fig:autocorr}  
\end{figure}

The conjectures of the previous section  have to be confirmed by a FSS analysis.
As well known in Monte Carlo simulations, the energy auto-correlation time 
$\tau_E$ is a good test to know about the order of the transition. In 
Table~\ref{tab:table2}, 
we have given the characteristic values of $\tau_E$ for different sequences 
and sizes at
the critical point. The number of MC iterations is always of order $10^4\tau_E$ to
ensure reliable results. The autocorrelation time is shown in 
Fig.~\ref{fig:autocorr} in a semi-logarithmic scale.

The numerical data for PF and TF sequences can be fitted by a power-law $\tau_E\sim
L^z$, with a very small dynamical exponent presumably linked to a logarithmic
behaviour, while in the case of TM, $\tau_E$ is exponentially diverging 
$\tau_E\sim L^{d/2}{\rm e}^{2\sigma L^{d-1}}$ where $\sigma$ is an order-disorder
interface tension.
These results support strong evidences that in the case of Paper-Folding and Three-Folding
sequences, the fluctuations are strong enough to soften the transition to
a second-order regime. In the case of Thue-Morse, even if the autocorrelation 
time,
compared to the periodic case, is lowered
by the fluctuations, the transition remains first-order. We note that the 
data corresponding
to the PS were not fitted since one has to go to strongly first order 
transitions ($q=15$) to observe the predicted exponential 
behaviour in the pure system~\cite{Gupta}.

\subsection{Bulk properties}

We can now enter upon a more refined characterization of the phase transition in
the two regimes. Since the critical point is exactly known, 
FSS techniques are well indicated  to get  accurate 
results. We made different series of simulations with free-free ($f|f$),
fixed-fixed ($F|F$), and
mixed ($f|F$) or ($F|f$) boundary 
conditions (BC)  in the horizontal direction. From all simulations
we can extract the singular behaviour of the magnetization and of the 
susceptibility, since there is no regular contribution and, at the critical
point we can write
\begin{equation}
M(K_c,L)={\cal A}_ML^{-\beta/\nu},
\label{eq:M}
\end{equation}
\begin{equation}
\chi(K_c,L)={\cal A}_\chi L^{\gamma/\nu},
\label{eq:Chi}
\end{equation}
where ${\cal A}_X$ are non-universal critical amplitudes. 
On the other hand, 
the four series of simulations are necessary in order to extract the singularity
 associated to the energy
 density which contains a regular part including 
 both a bulk $E^{(0)}$ and a surface $E^{(-1)}
 \times L^{-1}$ 
 contribution~\cite{BercheBercheTurban96,KarevskiLajkoTurban97}. This latter
 part must be
 split in two terms, since the two surfaces are different. We thus 
 have~\footnote[2]{We note the the singular surface terms 
 being less divergent, they 
 only add a correction to scaling.}
\begin{eqnarray}
E_{(l|r)}(K_c,L)&=&E^{(0)}(K_c,L)\nonumber\\
&+&(E^{(-1)}_l(K_c,L)+E^{(-1)}_r(K_c,L))\times L^{-1}\nonumber\\
&+&
{\cal A}_{(l|r)}L^{(\alpha-1)/\nu}+\dots
\label{eq:E}
\end{eqnarray}
where $(l|r)$ specifies the BC's (free or Fixed) for the left and right surfaces.
We note that our simulations being performed in a cylinder geometry, the Euler
number  vanishes, and thus there is no $\ln L$ term in the free 
energy~\cite{CardyPeschel88,Cardy88}.
The asymptotic values of the bulk energy density with different BC's are the 
same in the 
thermodynamic limit, but the 
amplitudes of the finite-size corrections being different,
the regular contributions cancel in the combination:
 \begin{eqnarray}
\Delta E(K_c,L)&=&E_{(F|F)}+E_{(f|f)}-E_{(f|F)}-E_{(F|f)}\nonumber\\
&=&
[{\cal A}_{(F|F)}+{\cal A}_{(f|f)}-{\cal A}_{(f|F)}-{\cal A}_{(F|f)}]
L^{(\alpha-1)/\nu}\nonumber\\
\label{eq:DeltaE}
\end{eqnarray}
 leaving a pure power-law.

\begin{figure}[ht]
\epsfxsize=9.5cm
\begin{center}
\mbox{\epsfbox{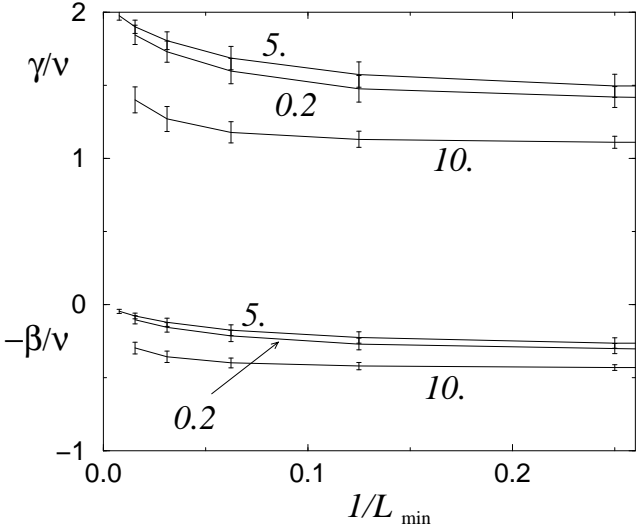}}\end{center}
\caption{Effective size-dependent exponents associated to  the susceptibility
and magnetization
in the case of TM sequence for different values $r=K_1/K_0=0.2$, 5, and 10, of the
coupling ratio (Note that $r=0.2$ corresponds to a perturbation of the same
strength than $r=5$). The error bars correspond to the standard deviations of the 
corresponding power-law fits.}
\label{fig:cross_TM}  
\end{figure}

The scaling dimensions can be deduced from log-log plots of the different quantities
{\it vs.} the system size.
In the case of the Thue-Morse sequence, a  
crossover appears,
as the size increases,  towards a behaviour which resembles the periodic one, 
characterized
by a vanishing exponent for the magnetization. This effect is visible in 
Fig.~\ref{fig:cross_TM} where the evolution of size-dependent effective 
exponents is shown. Similarly, after the crossover
regime,  the susceptibility
is described by an exponent $\gamma/\nu$ close to the value $d=2$.
The behaviour of the energy density difference exhibits  
log-periodic oscillations, characteristic of systems with
discrete scale invariance. It makes more difficult a 
quantitative analysis,\footnote[3]{The discrete rescaling factors for 
the different sequences take the values: 4 (TM), 2 (PF), and 3 (TF).
} 
but a tendency to a decreasing
slope for the largest size is nevertheless observed.

Certainly,  a careful analysis is needed to avoid the crossover effects from
small sizes to the true fixed point behaviour in the infinite lattice size
limit, so we 
 define the following procedure:
 From the log-log curves between $pL_{\rm min}$ and $pL_{\rm max}$, one
  determines 
an effective
exponent $x(L_{\rm min})$ for each quantity; 
then the smaller size is canceled from the data 
and the whole procedure is 
repeated until only the three or four largest sizes  remain. The effective exponent is 
then
plotted against $L_{\rm min}^{-1}$. This prescription makes appearent the
crossover effects and enables us to identify
unambiguously the asymptotic regime\footnote[4]{One can nevertheless mention that the smallest strip size 
is not of great value, since it is smaller than the correlation length of the pure model
at the transition point in the disordered phase~\cite{BuffenoirWallon93}.}. 
Different values of the 
aperiodic perturbation ($r=0.2$, 5, and 10) are shown in 
Fig.~\ref{fig:cross_TM} in the case
of TM sequence. At small sizes, the system is still strongly under the 
influence of the fluctuations induced by the aperiodic distribution of
couplings, while as the size increases, the effective exponents converge
towards trivial values which are characteristic of a first-order regime.
The ratio $\beta/\nu=0$ is indeed characteristic of a discontinuity of the
order parameter, while $\gamma/\nu=2$ is consistent with this discontinuity and with
the scaling law $2\beta/\nu+\gamma/\nu=d$.
This behaviour of effective exponents is the signature that the aperiodic 
fluctuations are eventually irrelevant.
This is corroborated by the fact that the crossover takes place at larger
sizes when the perturbation amplitude becomes stronger ($r=10$).

On the contrary, the two other aperiodic sequences 
exhibit power-law behaviours with non-trivial exponents.
Since a second-order
phase transition occurs for these sequences, the question of the stability
  of the new fixed point has to be considered. A numerical study at 
  different ratios of interactions\footnote[5]{The simulations are more
complete in the case $r=5$, since we
have one size less for the other values.} $r=0.2$, 2, 3, 
5, and 10  shows that the corresponding exponents remain stable for strong enough
perturbations  ($r$ not too  close to the pure system value 
1).

\begin{figure}[ht]
\epsfxsize=9.5cm
\begin{center}
\mbox{\epsfbox{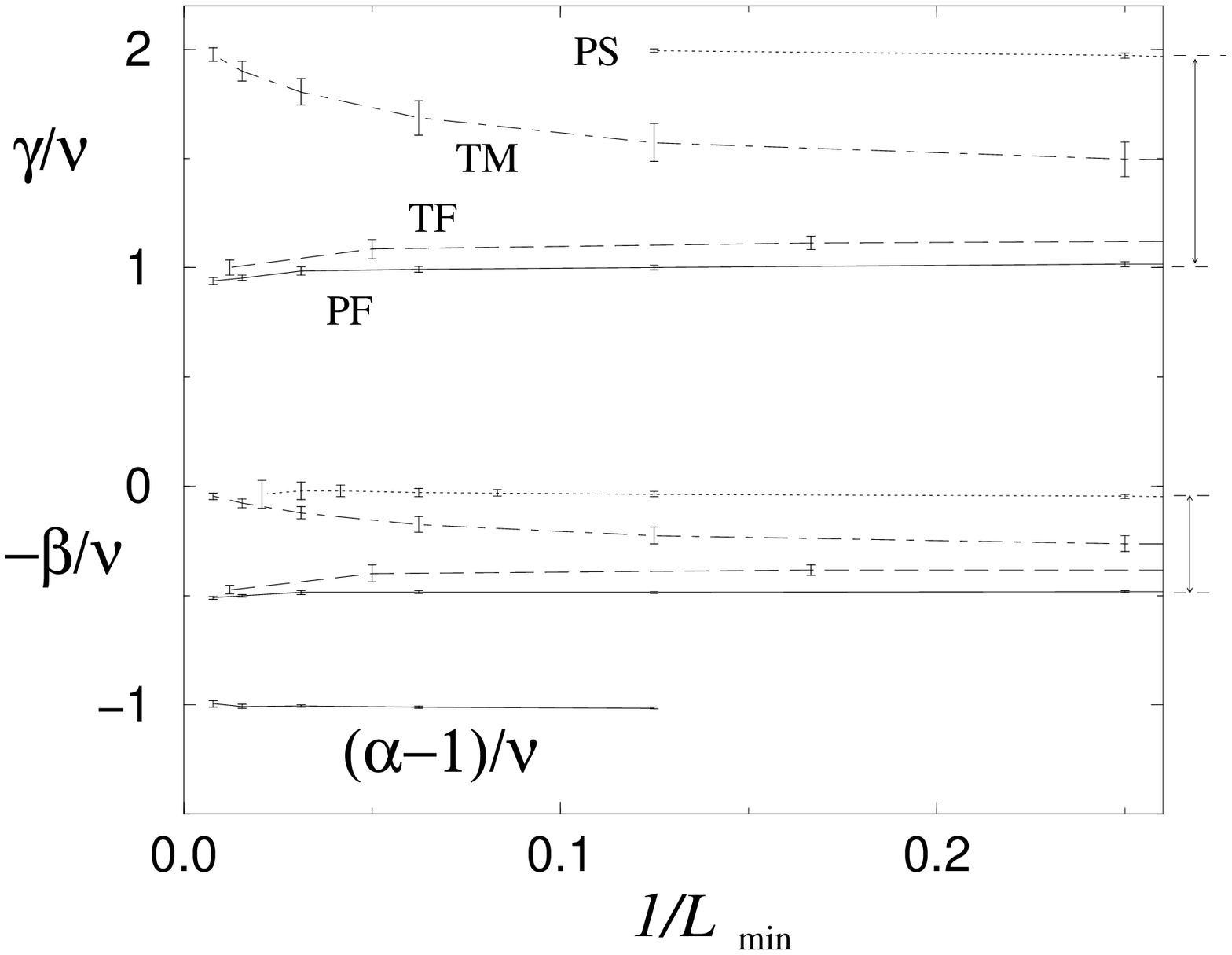}}\end{center}
\caption{Effective size-dependent exponents associated to the susceptibility,
magnetization, and energy density difference (dotted line: PS, dot-dashed line: TM,
long-dashed line: TF, and solid line: PF) for $r=5$. Arrows on the right of the
figure demarcate the curves corresponding to $\gamma/\nu$ and $-\beta/\nu$.}
\label{fig:x_M-Chi_DE}  
\end{figure}

The numerical results of power-law fits
in the linear part of log-log plots 
are given in Table~\ref{tab:table3} for PS, TM, PF and TF 
sequences.
\begin{table}
\small
\caption{Bulk critical exponents obtained by the slope of 
finite-size
scaling results   for the fours sequences. The numbers in parentheses 
give the 
estimated uncertainty in the last digit.}
\vglue0mm\begin{center}
\begin{tabular}{@{}*{6}{l}}
\br
 & \centre{2}{\rule{0mm}{1mm}PS} & \centre{2}{TM} \\
 & \crule{2} & \crule{2} \\
 & $\beta/\nu $ & $\gamma/\nu $  & $\beta/\nu $ & $\gamma/\nu $  \\
\hline
$r=0.2$	&0.046(6)  &2.01(1)  &0.10(3)  &1.81(7)  \\
$r=5.$	&0.072(4)  &1.97(1)  &0.08(2)  &1.90(5)  \\
$r=10.$	&0.045(2)  &  2.02(1)& cross.$^a$        &cross.         \\
\br
 & \centre{2}{\rule{0mm}{1mm}PF} & \centre{2}{TF} \\
 & \crule{2} & \crule{2} \\
 & $\beta/\nu $ & $\gamma/\nu $ & $\beta/\nu $ & $\gamma/\nu $  \\
\hline
$r=2.$	&0.464(7)  &1.13(2)  &--$^b$  &--   \\
$r=3.$  &0.475(8)  &1.07(2)  &--  &--   \\
$r=5.$  &0.480(3) &1.017(9) &0.43(2)  &1.15(2)   \\
$r=10.$ &0.49(1)  &1.01(1)  &0.44(2)  &1.09(3)   \\
$r=0.2$ &0.46(1)  &1.04(2)  &0.43(1)  &1.15(2)   \\
\br
 & \centre{4}{\rule{0mm}{1mm}$(1-\alpha)/\nu$}\\
& \crule{2} & \crule{2} \\ 
& PS & TM & PF & TF \\
 \hline
$r=5.$  &0.05(4) &osc.$^c$ &1.015(4)  &osc.   \\
\br
\end{tabular}
\\
\medskip\end{center}
{\footnotesize $^a$ ``cross.'' means that the 
crossover is still too strong to allow any linear regime in the
log-log plots.\hfill\break
$^b$ The symbol -- means that the corresponding
runs have not been performed. \hfill\break
$^c$ ``osc.'' means that the 
log-periodic
oscillating behaviour does not allow any precise estimation of the exponent.} 
\label{tab:table3}
\end{table}
Since the values deduced from these power-law fits seem to remain
stable with respect to the perturbation amplitude, we can use the effective 
exponents (Fig.~\ref{fig:x_M-Chi_DE}) to determine a more accurate value of
the critical exponents 
deduced from the  extrapolation at infinite size. 
The results are given in Table~\ref{tab:table4} for
all sequences at $r=5$, for which value we have the more exhaustive 
numerical results. Log-periodic oscillations again appear in the behaviour of 
the 
energy density combination in the TF sequence, but the tendency is coherent 
with
the behaviour of PF sequence.

\begin{table}
\small
\caption{Bulk critical exponents obtained by extrapolation
at infinite size of 
finite-size
scaling results for the sequences PS, TM, PF and TF with a ratio $r=5$. 
}
\vglue0mm\begin{center}
\begin{tabular}{@{}*{6}{l}}
\br 
 &\rule{0mm}{1mm}  PS & TM & PF & TF \\
\hline
$\beta/\nu$ 	&  $0.012$    & $0.020$ & $0.499$ & $0.508$   \\
$\gamma/\nu$	&  $1.986$    & $1.993$ & $0.995$ & $1.009$   \\
$(1-\alpha)/\nu$&  $0.05 $    &      osc.$^a$ & $1.001$ & osc.   \\
\br
\end{tabular}
\\
\medskip\end{center}
{\footnotesize $^a$ ``osc.'' means that the 
log-periodic
oscillating behaviour does not allow any precise estimation of the exponent.}
\label{tab:table4}
\end{table}

From these values of the critical exponents, we can deduce the scaling 
dimensions associated to the temperature and the magnetic field, 
$y_t=d-\frac{1-\alpha}{\nu}\approx 1.00$
and $y_h=d-\frac{\beta}{\nu}=\frac{d+\gamma/\nu}{2}\approx 1.50$ at the new fixed point
for PF and TF. It could be  surprising to obtain, within the 
precision of our results,
the same fixed point for these aperiodic perturbations, but we can
mention here that both of them have the same wandering exponent $\omega=0$.

\subsection{Surface properties}

\begin{figure}[ht]
\epsfxsize=9.5cm
\begin{center}
\mbox{\epsfbox{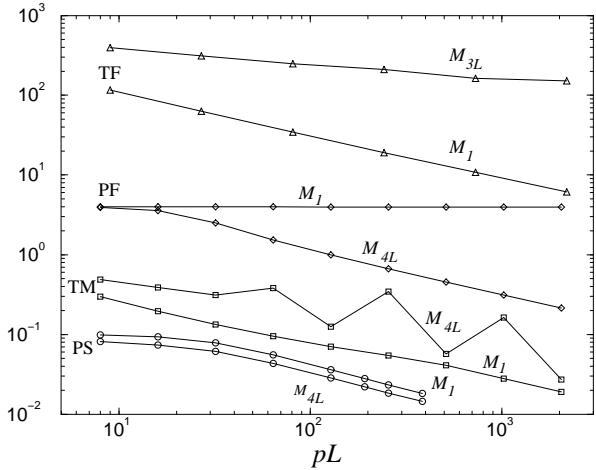}}\end{center}
\caption{Local surface magnetization at both ends of the system for the four sequences
considered in the text. The curves have been shifted for clarity.}
\label{fig:M_surf}  
\end{figure}

The surface properties can also be investigated, and here we ask if
the aperiodic perturbations are also liable to modify the surface critical 
behaviour.
We determined numerically the value of the order parameter at both surfaces
$j=1$ and $j=pL$ for the four sequences considered. It gives the corresponding
exponents, called $\beta_1/\nu$ and $\beta_{pL}/\nu$, respectively. The
log-log plots are shown in Fig.~\ref{fig:M_surf} and the exponents, deduced from the slopes
of log-log plots in the linear regime, are given in Table~\ref{tab:table5}.

\begin{table}
\footnotesize\small
\caption{Surface critical exponents$^a$ obtained by the slope of 
finite-size
scaling results   for the four sequences.}
\vglue0mm\begin{center}
\begin{tabular}{@{}*{6}{l}}
\br 
& \centre{2}{\rule{0mm}{1mm}PS} & \centre{2}{TM} \\
&  \crule{2} & \crule{2} \\
&  $\beta_1/\nu $ &  $\beta_{4L}/\nu$ &$\beta_1/\nu $ &  $\beta_{4L}/\nu$ \\
\hline
$r=5.$	&0.60(1) &0.59(1) &0.55(2) &osc.$^b$ \\
$r=10.$	&0.60(1) &0.60(1) &cross.$^c$ &cross.\\
$r=0.2$	&0.56(1) &0.61(1) &0.60(2) &osc.\\
\br
& \centre{2}{PF} & \centre{2}{TF} \\
&\crule{2} &\crule{2}  \\
  &$\beta_1/\nu $ 
 &  $\beta_{4L}/\nu$ &$\beta_1/\nu $ &  $\beta_{3L}/\nu$\\
\hline
$r=2.$	&0.013(3) &0.58(1)&--$^d$ &--  \\
$r=3.$	&0.0023(6) &0.58(1) &-- &--  \\
$r=5.$	&0.0000(0) &0.58(1) &0.530(4) &0.17(1)  \\
$r=10.$	&0.0000(0) &0.58(1) &0.532(2) &0.095(3)  \\
$r=0.2$	&0.54(1) &0.000(0) &0.19(1) &0.61(2)  \\
\br
\end{tabular}\\
\medskip\end{center}
{\footnotesize $^a$The numbers in parentheses give the estimated 
uncertainty in the last digit.\hfill\break
$^b$ ``osc.'' means that the 
log-periodic
oscillating behaviour does not allow any precise estimation of the exponent.\hfill\break
$^c$ ``cross.'' means that the 
crossover is still too strong to allow any linear regime in the
log-log plots.\hfill\break
$^d$ The symbol -- means that the corresponding
runs have not been performed. \hfill\break}
\label{tab:table5}
\end{table}

\begin{figure}[ht]
\epsfxsize=9.5cm
\begin{center}
\mbox{\epsfbox{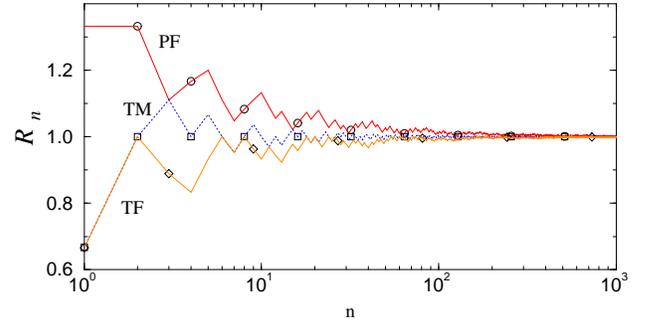}}\end{center}
\caption{Average coupling at a length scale $n$ from the left surface for the sequences PF, 
TM and TF ($r=2$). The interactions are enhanced, on average, 
close to the surface for PF, while
they are reduced for TF, and remain constant in the case of TM.}
\label{fig:fluc}  
\end{figure}

We can observe that the PS and TM systems exhibit analogous singularities at
the surfaces. It is known in the pure case that, even with a first-order 
phase transition in
the bulk of the system, a second-order phase transition is obtained at the 
surface. The critical exponents keep constant values around 0.6 independently
of the interaction ratio $r$.
In the case of PF and TF sequences, a second-order regime which depends on the
coupling ratio is obtained. The transition is strenghtened when the coupling  are sufficiently
enhanced, and one can even observe a first-order transition (in the
case of PF) at the surface.\footnote[5]{This first-order transition is only possible here because the system
being two-dimensional, the surface alone cannot order above the 
critical temperature
where the bulk is not ordered. In higher dimensions, this regime should 
lead to a
surface transition.}
This result can be understood by the behaviour of the average coupling 
at a length scale $n$ in the vicinity
of the left surface for example, 
$\bar K_n$, compared to the asymptotic average coupling $\bar K_\infty$:
\begin{equation}
R_n=\frac{\bar K_n}{\bar K_\infty}=\frac{\rho_n(r-1)+1}{\rho_\infty(r-1)+1},
\quad\rho_n=\frac{1}{n}\sum_{k=1}^n f_k.
\label{eq:Kav}
\end{equation}
This ratio is always greater than 1 when $r>1$ for PF and produces 
a significative enhancement of the interactions
close to the boundary, leading to a decrease of the exponent of the surface
magnetization. The contrary 
happens with TF as shown in Fig.~\ref{fig:fluc}.
We have also checked that these exponents are characteristic not only of the
local boundary behaviour, but also of an average surface property, since the
average of the order parameter over a few rows (from 2 to 10) 
in the vicinity of the surface reproduces the same exponents.

\subsection{Crossover effect or marginal variation of the exponents for
PF and TF sequences?}

The question one poses in this section is whether the small observed variations 
of the 
critical exponents in the second-order phase transition regime results from
a crossover effect or from a marginal behaviour.
It is interesting here to make a comparison with what occurs in the Ising model
case $q=2$. For this model, the PF and TF sequences are known to lead to a marginal
behaviour with continuoulsy varying critical exponents. 
The surface properties of this system have 
been intensively 
investigated~\cite{BercheBercheTurban96,IgloiTurbanKarevskiSzalma97}, but the 
coupling distribution used here being different from the one used in  
previous works,  
we cannot directly compare the values of the exponents. 
We performed a few simulations for $q=2$, and the results are given in Table~\ref{tab:table6}. It 
appears
clearly that the values of the different critical exponents exhibit, as 
expected, 
a continuous
variation with the amplitude of the interactions.  The aim of 
these supplementary
simulations is to show unambiguously that a marginal variation of the 
exponents
is strong enough to be distinguished numerically from the crossover effect 
at small perturbations. 
Our results suggest that the small variation observed in the Potts 
model case is probably due to this
latter situation.
\begin{table}
\small
\caption{Bulk and surface critical exponents$^a$ obtained by the slope of 
finite-size
scaling results   for PF sequences in the case of the Ising model $q=2$.}
\vglue0mm\begin{center}
\begin{tabular}{@{}*{6}{l}}
\br 
   $r$ & $\beta/\nu$   & $\gamma/\nu$  &  $\beta_1/\nu$  &  $\beta_{4L}/\nu$    \\
\hline
0.2 &0.337(2)&1.37(1)&0.497(1)&0.0203(4)\\
5   & 0.360(2) & 1.351(9)   & 0.0176(7) & 0.517(3)  \\
10 &0.407(1)&1.233(6)&0.0044(2)&0.525(4)\\
\br
\end{tabular}
\\
\medskip\end{center}
{\footnotesize $^a$The numbers in parentheses give the estimated uncertainty in the last digit.
}\label{tab:table6}
\end{table}


\section{Conclusion}

In this paper, we have shown that the deterministic fluctuations generated by an 
aperiodic modulation
of nearest-neighbour couplings in the eight-state Potts model 
produce a softening
of the 
transition, and are even liable to induce a second-order phase transition. 
It 
happens when the fluctuations around the average coupling are strong enough, 
and it is the
case for the Paper-Folding and Three-Folding sequences, although they are 
characterized by a vanishing
wandering exponent. In the case of the Thue-Morse sequence, the wandering exponent
being $-\infty$, the transition remains of first-order, and the scaling
dimensions keep their pure values. These results are consistent with Luck's 
criterion~\cite{Luck93}, provided that we replace the correlation-length exponent $\nu$ by its
trivial value $1/y_t=1/d$ at the first-order fixed point. The crossover exponent associated
to the aperiodic distribution $\Phi=1+(\omega-1)/d$ is then positive (relevant
perturbation) for PF and TF sequences
while it is negative (irrelevant) in the case of TM sequence.

\begin{figure}[ht]
\epsfxsize=8cm
\begin{center}
\mbox{\epsfbox{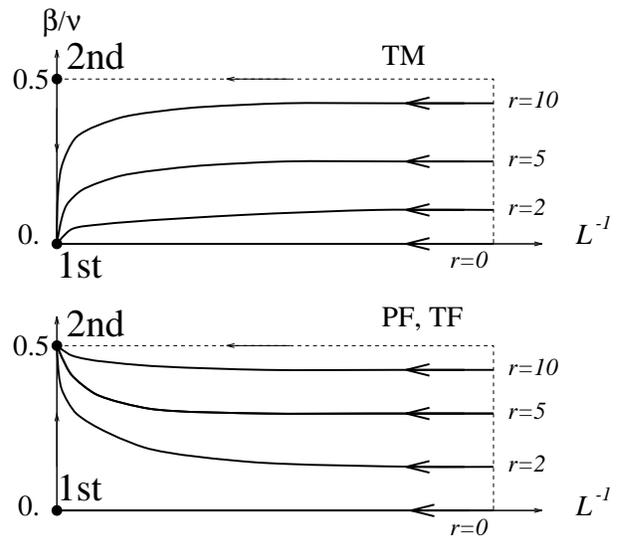}}\end{center}
\caption{Evolution of the size-dependent effecive exponents 
for the two types of aperiodic sequences.}
\label{fig:RG}  
\end{figure}

The analysis of the bulk properties shows that the new fixed point 
exponents (PF and TF)  are stable, i.e. do not depend, up to small 
crossover effects, on the
value of the perturbation amplitude. This is clearly different from the 
marginal behaviour encountered for similar sequences in the Ising model case.
We can furthermore notice that the new universality class 
seems to be robust, i.e. the same for
both sequences, a result which is not a priori obvious. 
One can nevertheless mention that our results are coherent with the stability
of the new fixed point, which requires a non-positive value for 
the crossover exponent
$\Phi'$ in the second-order regime. 
With $\omega=0$, $\Phi'$ takes the value $\Phi'=1-\nu$
at the new fixed point. Using hyperscaling relation $\alpha=2-d\nu$ (which 
should hold, unless
anisotropic behaviour is found), the value $\alpha\simeq 0$,  obtained 
numerically for
PF, leads to $\nu\simeq 1$ and thus $\Phi'\simeq 0$.
It seems reasonable
to propose a renormalization group sketch, illustrated by the evolution of
the effective exponents, where the two possible sets
of exponents should correspond to two different  fixed points, the stability 
of which
depend on the strength of fluctuations (i.e. the value of the wandering 
exponent). The two types of situations are shown in 
Fig~\ref{fig:RG}. The periodic system, as it can be seen in 
Fig~\ref{fig:x_M-Chi_DE}, is not  influenced by 
the existence of the second fixed point.

The surface 
magnetization has also been analyzed and analogous conclusions can be given.
It is interesting to notice that a first-order surface transition (with a
second-order regime in the bulk) can be induced in the case of PF, i.e. the
exact contrary of the pure model behaviour.

There are still some open questions concerning aperiodic perturbations in these
systems. A possible anisotropic scaling behaviour could be 
obtained in the second-order
induced regime, since it occurs in the Ising model already. One should also 
investigate aperiodic sequences with diverging fluctuations characterized by
a positive wandering exponent, like Rudin-Shapiro for example. A second-order
transition should also be obtained, but its universality class could be 
different.

\begin{acknowledgement}
We gratefully acknowledge Lo\"\i c Turban for stimulating discussions, and 
PE B thanks Prof. Kurt Binder for
hospitality in Mainz. We also thank the referee for his/her constructive criticisms.
This
work was supported by CNIMAT under project No 155C98 and by the Centre 
Charles Hermite. 
\end{acknowledgement}


\end{document}